\documentclass[openacc]{rstransa}%%%%where rstrans is the template name

\titlehead{Research}

\begin{document}

%%%% Article title to be placed here
\title{The connection between solar coronal abundances and the underlying lower atmospheric properties}

%%%% Author details
\author{
Paola Testa$^{1}$, Juan Martinez-Sykora$^{2,3,4}$, Bart De Pontieu$^{3,4}$, Alberto Sainz Dalda$^{2,3}$, David Long$^{5}$, Deborah Baker$^{6}$, David H.\ Brooks$^{6,7}$} 

%%%%%%%%% Insert author address here
\address{$^{1}$Harvard-Smithsonian Center for Astrophysics, 60 Garden St., Cambridge, MA 02193, USA\\
$^{2}$SETI Institute, 339 Bernardo Ave, Suite 200, Mountain View, CA, 94043, United States\\
$^{3}$Lockheed Martin Solar and Astrophysics Laboratory, 3251 Hanover St, Palo Alto, CA 94304, USA\\
$^{4}$Rosseland Centre for Solar Physics, University of Oslo, P.O. Box 1029 Blindern, N-0315 Oslo, Norway\\
$^{5}$Centre for Astrophysics \& Relativity, School of Physical Sciences, Dublin City University\\
$^{6}$University College London, MSSL, Holmbury St. Mary, Dorking, Surrey, RH5 6NT, UK\\
$^{7}$Computational Physics, Inc., Springfield, VA 22151,USA
}

%%%% Subject entries to be placed here %%%%
\subject{astrophysics, stars, spectroscopy}

%%%% Keyword entries to be placed here %%%%

%%%% Insert corresponding author and its email address
\corres{Paola Testa\\ 
 \email{ptesta@cfa.harvard.edu}}

%%%% Abstract text to be placed here %%%%%%%%%%%%
\begin{abstract}
Elemental abundances in the solar corona and solar wind are often observed to differ from those in the solar photosphere, most commonly showing an enhancement of low first ionization-potential (FIP) elements (the FIP effect). The observational evidence of the connection between the chemical fractionation in the solar atmosphere with FIP suggests that the mechanisms responsible for this effect take place in the chromosphere, where low-FIP elements are mostly ionized, while high-FIP elements remain mostly neutral. 
We discuss the findings of recent observational studies that have investigated the possible footprint of coronal abundance anomalies in the lower atmosphere. We also discuss the limitations of current observations, and future perspectives on addressing this important open issue in solar physics.
\end{abstract}
%%%%%%%%%%%%%%%%%%%%%%%%%%%

\maketitle

%%%%%%%%%% Insert the texts which can accomdate on firstpage in the tag "fmtext" %%%%%

%\begin{fmtext}
%\end{fmtext}

\section{Introduction}
%%%% Insert A head here

The solar atmosphere spans from the visible surface, the photosphere ($\sim 5800$~K), through the dynamic chromosphere ($\sim 10,000$~K) and thin transition region ($\sim 10^5$~K), to the hot corona ($\gtrsim 10^6$~K). These different layers are shaped by the changing magnetic structure, from the convection dominated surface, through the steep temperature and density gradients of the interface to the magnetically controlled extended corona, which gives rise to the solar wind.
The outer solar atmosphere has been the object of intense study and fascination for scientists for centuries during eclipses, and much more in detail in the last few decades since observations from space have allowed access to its high energy (mainly X-rays and Extreme Ultraviolet) emission which constitutes a dominant portion of its overall emission \cite{Testa2023solar}.   The plasma in the solar atmosphere presents several intriguing properties that are not yet fully understood, including its high temperatures ($\gtrsim 1$~MK, orders of magnitude larger than at the surface), the dynamic nature of its emission often leading to large flares that significantly impact the heliosphere and Earth, and its chemical composition, which is highly structured and different from the photospheric makeup \cite{widing1992element,feldman1992elemental,vonSteiger2000,feldman2000,testa2015}.

Early spectroscopic observations of the solar atmosphere quickly established the presence of anomalies in the chemical composition of the coronal plasma compared to the known photospheric abundances \cite{Pottasch1963}. Element abundances at odds with photospheric composition were then also found in the solar wind \cite{Meyer1985}. The fractionation of the elements in the solar atmosphere has been observed to depend on the first ionization potential (FIP) of the element.
In particular, observations in much of the corona and slow solar wind typically show low-FIP elements (such as Fe, Mg, and Si) enhanced by factors $\sim$2-4 (so-called, "FIP bias") relative to their photospheric abundance, whereas coronal holes and fast wind generally show weak FIP bias \cite{Laming2015}. 
This dependence on the first ionization potential implies that the chemical fractionation process largely occurs in the solar chromosphere, where elements with low FIP are mostly ionized while high-FIP elements remain mostly neutral, and also suggests a connection between the fractionation mechanisms and the heating of the solar atmosphere (e.g., \cite{Testa2010,Laming2015,testa2015,Brooks2025,Baker2025,JMS2025}, and references therein).  

Spectroscopic studies, in X-rays and EUV bands, of solar-like stars show that chemical fractionation is a characteristic feature of stellar coronae, although with a broad range of different manifestations  (e.g., \cite{Testa2010,testa2015,Laming2015,Brooks2025} and references therein). In fact, stellar coronal composition shows a variety of behaviors ranging from a solar-like FIP effect for low-activity to moderately active stars (e.g., $\alpha$~Cen, \cite{Drake1997}), to an inverse FIP effect (IFIP), where high-FIP elements are enhanced in the corona (e.g.,\cite{Brinkman2001}), observed in most active stars (active binaries, M-dwarfs, young rapidly rotating stars,..). 
Possible dependencies on stellar parameters have emerged from studies of statistical samples of X-ray and EUV spectra of stars with a broad range of stellar parameters, including spectral type, stellar activity level, and age (e.g., \cite{Wood2010,Wood2013,Telleschi2005}). Overall, there is observational evidence that the coronal composition generally depends on the activity level and on the spectral type \cite{testa2015,Seli2022}. The widely observed coronal abundance anomalies indicate that the physical processes leading to the chemical fractionation in stellar outer atmospheres are a fundamental property of magnetically heated atmospheres \cite{testa2015,Laming2015}. Combined studies of solar and stellar coronal abundances therefore provide a powerful laboratory for constraining and testing fractionation models, and for understanding chromospheric and coronal physics across a wide range of magnetic activity and energy regimes.

The connection between solar wind in-situ measurement and remote sensing data also shows a possible pathway to trace some patches of solar wind to the sources at the Sun (e.g., \cite{Raymond1997,Brooks2011,Brooks2015,Abbo2016}). Therefore, studies of abundance anomalies can also help us link the solar atmosphere to solar wind and mass loss, which significantly affect the solar evolution and its impact on the interplanetary space and on planets. 

Solar and stellar observations indicate a reality much more complex than the relatively straightforward scenario depicted above. For instance, spatial and temporal abundance variations in the solar corona show more complicated patterns \cite{Brooks2025} with, for instance, some closed coronal structures, such as some regions close to sunspots, presenting very little fractionation or even an IFIP effect (e.g., \cite{{Doschek2015,Doschek2016,Baker2021,Baker2025}}). Additionally, in newly emerged flux regions, and in transient heating events, such as microflares and flares, the coronal plasma is typically characterized by abundances close to photospheric values (e.g., \cite{McKenzie1992,Widing1997,Young1997,Widing2001,Warren2016,Baker2018}). These results echo studies of stellar flares that suggest that during these intense transient heating events the coronal composition is observed to be closer to the stellar photospheric values (e.g.,\cite{Nordon2008,Kurihara2025}). Furthermore, during solar flare observations an inverse FIP effect is sometimes observed  \cite{Doschek2015,Doschek2016,Brooks2018,Baker2019,Baker2020,To2021}. 
Spatially and temporally resolved solar observations have also shown highly localized and variable abundance anomalies during flares with a FIP effect at the looptop and IFIP at the footpoints of flaring structures \cite{Doschek2018,To2024}. These observations offer a unique probe of fractionation mechanisms, and provide some of the clearest observational evidence that these processes are dynamic and chromospheric in origin.

Recent models provide a theoretical framework to interpret the observed abundance anomalies.
The physical process generally embraced as the most likely cause of the chemical fractionation in the solar and stellar atmospheres is the "ponderomotive" force, which arises from spatial gradients in the wave electric field energy and acts on ions and electrons, but not on neutrals, thus providing
a mechanism of ion-neutral separation (other fractionation mechanisms, generally unable to explain the wide range of observed FIP/IFIP effects, have been discussed in e.g., \cite{Drake2003,Laming2009,Testa2010,Bochsler2000} and references therein). 
In a semi-analytical model developed by Laming (e.g., \cite{Laming2015,Laming2025}) widely used for the interpretation of observations, the ponderomotive force arises from the strong reflection and refraction of Alfv{\'e}nic and fast-mode MHD waves at the steep gradients in density and Alfv{\'e}n speed near the upper chromosphere/transition region boundary. Recently, time-dependent multi-fluid MHD simulations are being developed \cite{MartinezSykora2023}, to self-consistently generate waves, currents, and ion-neutral drifts from photospheric driving and magnetic evolution (see more in Sec.~\ref{fip_chromo}\ref{limitations}).

The examples discussed so far clearly demonstrate how investigating the chemical fractionation in the atmosphere of the Sun and other stars provides a unique diagnostic of fundamental processes difficult to observe directly such as the heating of stellar atmospheres, and ion-neutral interactions occurring in the chromosphere.
The connection between the chromosphere and the upper atmospheric layers is definitely key to understanding the chemical fractionation in the solar atmosphere and to constraining models. 
In this paper we will concisely review the results of observational studies exploring the possible relationship between coronal FIP bias and chromospheric conditions.  In particular, here we focus on results of analysis of coordinated on disk observations with the Hinode/Extreme-ultraviolet Imaging Spectrograph (EIS \cite{Culhane2007}, which measures coronal FIP bias) and the Interface Region Imaging Spectrograph (IRIS \cite{DePontieu2014}, which provides unique chromospheric spectral observations) (Sec.~\ref{fip_chromo}). We also briefly discuss the limitations of current solar observations and of existing models (Sec.~\ref{fip_chromo}\ref{limitations}). We draw some conclusions and discuss future prospects in Sec.~\ref{conclusions}.

\section{Connection between coronal abundance anomalies and chromospheric properties}
\label{fip_chromo}
%\subsection{}
%%%% Insert B head here

Despite the importance of the link between the chromospheric properties and the coronal abundance anomalies, only a very limited number of observational studies aim at addressing this issue.
There have been a set of studies investigating the propagation of Alfv{\'e}n waves in the solar chromosphere, using spectropolarimetric inversion of ground based observations of areas with different observed coronal FIP bias \cite{Baker2021,Murabito2021,Stangalini2021,Murabito2024,Lee2025}, to explore the wave connection to coronal chemical fractionation. This topic is however discussed in detail in another articles in this collection \cite{Stangalini2025,Laming2025}, therefore, here, we focus on reviewing the results of investigations that have analyzed coordinated UV spectral observations of the corona (with Hinode/EIS) and of the chromosphere and transition region (with IRIS), to find any evidence of the footprints of the FIP effect in the lower atmosphere. IRIS provides observations at unprecedented combined spatial resolution, high cadence, and high spectral resolution in key chromospheric and transition region lines (e.g., Mg\,{\sc ii}, C\,{\sc ii}, Si\,{\sc iv}, O\,{\sc iv}), with simultaneous slit-jaw imaging and tight coordination with Hinode and SDO.

\begin{figure}[!h]
\centering\includegraphics[width=5.5in]{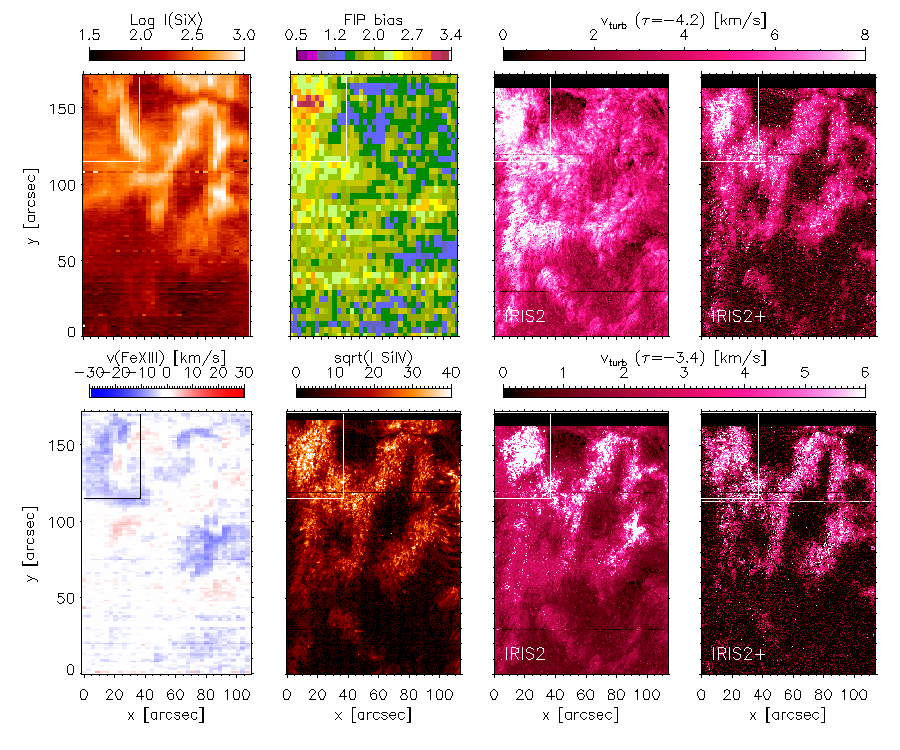}
%%% where xxxxxx name represents "figurename.eps"
\caption{Coronal abundances and lower atmospheric conditions from Hinode/EIS and IRIS coordinated observations of AR 12738 (adapted from Figure~5 of \cite{Testa2023}): Si\,{\sc x} (~1MK) intensity (top left), Fe\,{\sc xiii}\ Doppler velocity (bottom left), FIP bias (top row, second panel), Si\,{\sc iv} intensity (bottom row, second panel); the third and fourth columns show the chromospheric mictroturbulence derived from IRIS$^2$ and IRIS$^{2+}$ respectively at $\log\tau=-4.2$ (top) and $\log\tau=-3.4$ (bottom). The outflow region (marked by the box, $x \lesssim 20$; $y \gtrsim 120$) is characterized by enhanced mictroturbulence especially around $\log\tau=-4.2$. We note that the $\log\tau$ values for which we are showing the microturbulence maps fall within a range of $\log\tau$ where the inversions are better constrained (see \cite{SainzDalda2019,SainzDalda2024,SainzDalda2026} for details). The FIP bias map is derived by using an L1-norm inversion method \cite{MartinezSykora2026}, as described in \cite{Testa2023}.}
\label{fig_2019}
\end{figure}

While there is a wealth of observational studies of solar coronal abundance anomalies, principally using EUV spectral observations (e.g., \cite{Baker2015,Brooks2015,Mihailescu2022}, see also \cite{Laming2015} and \cite{Brooks2025} and references therein), the studies of \cite{Testa2023} and \cite{Long2024} have used coordinated IRIS observations to study the underlying chromospheric and transition region properties. \cite{Testa2023} used the Si/S abundance diagnostic (Si\,{\sc x} 285.38\AA, S\,{\sc x} 264.23\AA) to investigate the spatial and temporal distribution of the FIP bias in an active region over about 10 days, and IRIS spectra to derive the non-thermal broadening of the Si\,{\sc iv}\ transition region line, and the chromospheric turbulence from IRIS$^2$ inversions \cite{SainzDalda2019} of the Mg\,{\sc ii} h\&k lines.
They found that the FIP bias exhibits strong spatial structuring, with enhanced values in active-region outflows, moss, and some sunspot-adjacent regions, while the active-region core shows relatively low (but $>1$) and stable FIP bias. Temporal variability was generally modest, but more pronounced in moss regions, suggesting that fractionation in the moss might be linked to localized and evolving lower-atmosphere conditions rather than global active-region evolution.
The study identified enhanced chromospheric turbulence as well as nonthermal broadening in transition region lines for regions that map to high coronal FIP bias. Although the correlations are modest, this provides observational evidence that abundance anomalies have measurable “footprints” in the chromosphere. 
In Figure~\ref{fig_2019} we reproduce some of the results presented in \cite{Testa2023}, highlighting some correlation between the FIP effect (top row, second panel) and the chromospheric microturbulence (at $\log\tau=-4.2$ \footnote{We use $\log(\tau)$ to refer to $\log_{10}(\tau_{5000})$.}) derived from IRIS$^2$ inversions (third column), in the AR outflow region (at top left corner of the field of view).
The IRIS$^2$ inversions are based on machine and deep learning techniques to infer the thermodynamic conditions of the lower atmosphere from IRIS spectra by taking into account non-LTE conditions in the chromosphere \cite{SainzDalda2019}. The initial version of IRIS$^2$, applied in \cite{Testa2023}, uses the Mg\,{\sc ii} lines; a new version, IRIS$^{2+}$ \cite{SainzDalda2024,SainzDalda2026} expanded the inversion method to make use of constraints from more chromospheric lines, like C\,{\sc ii}, and resolved some of the possible ambiguities potentially affecting IRIS$^2$.  In particular, the inclusion of both the Mg\,{\sc ii} and C\,{\sc ii} lines in IRIS$^{2+}$ helps resolve the ambiguity between temperature and microturbulence in the chromosphere, both of which can lead to line broadening. As a result, the microturbulence derived from IRIS$^{2+}$ is expected to be more accurate than that of IRIS$^{2}$. Here we have therefore re-analyzed the datasets of \cite{Testa2023} with IRIS$^{2+}$ inversions, as a sanity check, and find that the enhancement of microturbulence at $\log \tau=-4.2$ appears even more confined to the outflow region than in the IRIS$^{2}$ results. 
In \cite{Testa2023} we showed the microturbulence at this $\log \tau$ value ($-4.2$) because, within the range of $\log\tau$ where the inversions are better constrained (see \cite{SainzDalda2019,SainzDalda2024,SainzDalda2026} for details), it was around that value that we observed a stronger correlation with FIP bias.
We note that the uncertainties of the specific connectivity between the high-FIP-bias coronal regions and the chromospheric footpoints, could certainly affect the observed correlation which might be underestimated. Additionally, the lack of strict simultaneity between the coronal and chromospheric data might also lead to a weak correlation pixel by pixel, even though a correlation is present in general in the high-FIP bias regions. Furthermore, FIP bias is generally found to be slowly evolving, and reflects the time integrated history of the chromospheric conditions, while the chromosphere dynamics is characterized by very short timescales, so this could also contribute to a poor instantaneous correlation, even in simultaneous observations.

%\vspace*{-2pt}

%\vspace*{-2pt}

\begin{figure}[!h]
\centering\includegraphics[width=5.5in]{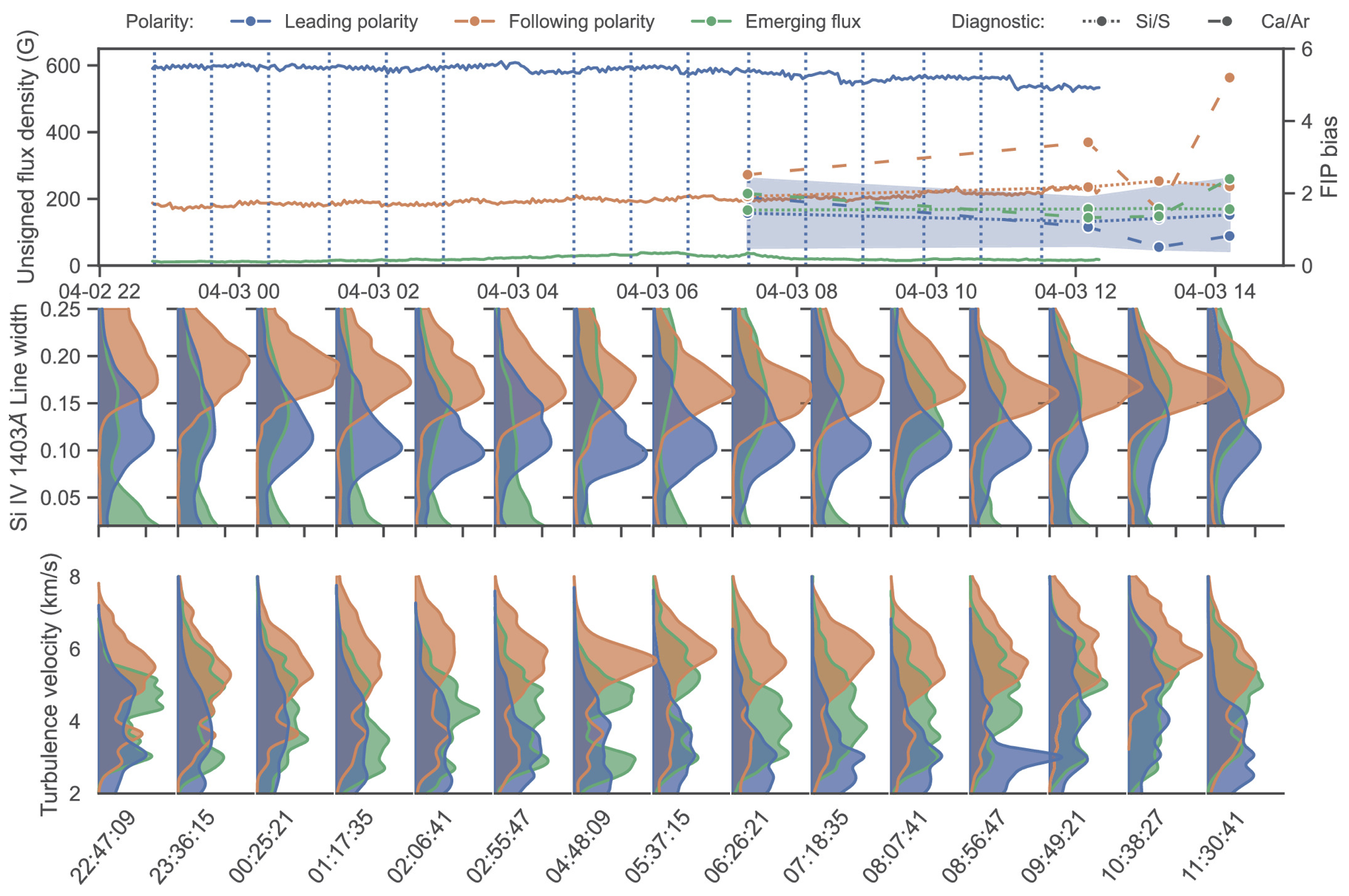}
%%% where xxxxxx name represents "figurename.eps"
\caption{Temporal evolution, over half a day, of coronal abundances, magnetic photospheric properties, and lower atmospheric conditions, for selected subregions (identified by different colors) in AR 12759 (adapted from Figure~5 of \cite{Long2024}). Top panel: unsigned magnetic flux (from SDO/HMI; solid lines), and FIP bias (dotted/dashed lines). A qualitative measure of the uncertainty associated with the FIP bias measurement is visualized with the blue shaded region which shows the uncertainty associated with the Si X/S X ratio.
Middle panel: Si\,{\sc iv} line width distribution in the different subregions, at different times. Bottom panel: distribution of mictroturbulence at $\log\tau=-4.2$ (derived from IRIS$^2$) in the different subregions, at different times.
High FIP bias (e.g., orange) is associated with enhanced Si\,{\sc iv} width, and mictroturbulence at $\log\tau=-4.2$.}
\label{fig_long24}
\end{figure}

The work by \cite{Long2024} built on the approach by \cite{Testa2023} by examining how chromospheric signatures of FIP bias might vary with magnetic polarity and substructure, and by implementing the additional Ca/Ar FIP effect diagnostic, which probes a different coronal temperature regime ($\sim 3$~MK vs $\sim 1$~MK of the Si/S diagnostic). 
\cite{Long2024} analysis of AR 12759 over six days, reveals FIP bias maps with pronounced spatial variability, including regions of enhanced FIP bias and localized inverse-FIP-like signatures. They found the abundance patterns to be closely linked to magnetic polarity and active-region substructure.
The coordinated IRIS observations indicate systematic differences in chromospheric dynamics between regions associated with enhanced and reduced coronal FIP bias, with higher nonthermal broadening and wave-like signatures occurring preferentially beneath regions of strong fractionation \cite{Long2024}. 
We show an example in Figure~\ref{fig_long24}, where high FIP bias (top row) appears correlated with the Si\,{\sc iv} line broadening (middle row), and the chromospheric microturbulence at $\log\tau=-4.2$ as derived from IRIS$^2$ inversions (bottom row).

%\vspace*{-7pt}

%\vspace*{-5pt}

These studies establish a direct observational bridge between coronal abundance anomalies and chromospheric dynamics, providing observational evidence of some footprint of the FIP effect in the chromosphere, and placing strong new constraints on both active-region outflow models and theories of elemental fractionation. IRIS observations, via IRIS$^{2+}$ inversions, can also provide upper limits on the amplitude of high frequency waves, assuming that the derived microturbulence can be ascribed to waves.

To further explore the connection between active region outflows and chromospheric microturbulence, we have also analyzed another dataset with the method used in \cite{Testa2023} (and now presented and tested in detail in \cite{MartinezSykora2026}). We use coordinated Hinode/EIS and IRIS observations of AR~12713 from 2018-06-17.  \cite{Polito2020} had analyzed this dataset focusing on the chromospheric and transition region properties of the outflow region, and found that the Si\,{\sc iv}, C\,{\sc ii}, and Mg\,{\sc ii} lines are significantly different from non-outflow regions (they use AR moss as a comparison), indicating a clear counterpart of the overlying coronal outflows. For this Hinode/EIS dataset we derive the FIP bias using the Si\,{\sc x}/S\,{\sc x} diagnostic, and derive chromospheric microturbulence using IRIS$^2$ and IRIS$^{2+}$ inversions. Unfortunately, the S\,{\sc x} line is overall rather weak in the outflow region, hampering the determination of the FIP bias in most of the observed outflow area; where some FIP bias values are derived (e.g., at the eastern boundary of the outflow region; $x < 10$, $90 \lesssim y \lesssim 110$) they appear possibly compatible with high FIP bias, as often  observed in outflow regions (e.g., \cite{Brooks2015,Testa2023}).
Similar to the results for AR~12738 (Figure~\ref{fig_2019}), enhanced mictroturbulence at $\log\tau=-4.2$ is found also in this outflow region.

The IRIS and EIS coordinated observations are possibly suggestive of an enhancement of chromospheric mictroturbulence in high-FIP bias moss regions, however at a deeper column mass ($\log\tau=-3.4$) compared to the case of the outflows. This is an intriguing possibility, and to some extent compatible with the fact that in moss (which is the transition region of hot AR loops) the high pressure shifts the emission at a given temperature down at higher column mass. However, the data we have analyzed so far are not adequate to corroborate this hypothesis, with the local non-simultaneity of the IRIS and Hinode spectra and their very different spatial resolutions especially affecting this type of investigation.

%\vspace*{-7pt}

\begin{figure}[!h]
\centering\includegraphics[width=5.5in]{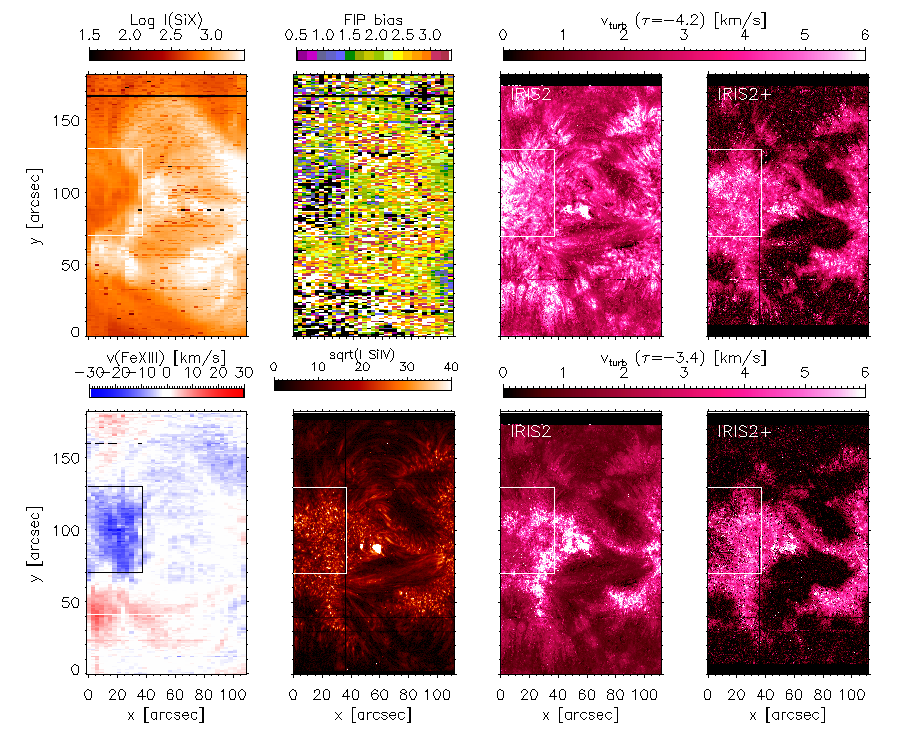}
%%% where xxxxxx name represents "figurename.eps"
\caption{Same as Figure~\ref{fig_2019} but for AR~12713 observed on 2018-06-17. For this Hinode/EIS dataset the lines used for the FIP bias diagnostic, especially S\,{\sc x}, are rather weak in the outflow region (to the E end of the shown field of view, as marked by the box), hampering the determination of the FIP bias in most of the observed outflow region. Like in the case of AR~12738, shown in Figure\ref{fig_2019}, enhanced mictroturbulence at $\log\tau=-4.2$ is found in the outflow region. The FIP bias map is derived by using an L1-norm inversion method \cite{MartinezSykora2026}, as described in \cite{Testa2023}.}
\label{fig_2018}
\end{figure}

%\vspace*{-5pt}

Despite the presence of some observational evidence of the footprints of the FIP effect in the chromosphere, as discussed here above, these analyses also present some puzzling results. First of all, the correlation between coronal FIP bias and transition region and chromospheric properties appears to be moderate (for instance, the correlation coefficient between FIP bias and microturbulence for the region of Figure~\ref{fig_2019} is $\lesssim 0.4$; although we should keep in mind the intrinsic differences in the data from the two instruments [see subsection~\ref{limitations}]).
Moreover, the observational relationship between coronal FIP bias, and chromospheric microturbulence and transition region non-thermal broadening appears rather complex. For instance, high-FIP bias areas close to sunspots do not seem to show a clear increase in microturbulence \cite{Testa2023,Long2024}.
The differing chromospheric turbulence signatures between high-FIP-bias outflow regions and those adjacent to the sunspot could be due to very different reasons. They may reflect distinct fractionation mechanisms operating in these environments, possibly also linked to the different magnetic properties in the two types of regions. They could alternatively indicate that the enhanced turbulence in outflows is associated with the generation of the flows themselves rather than being directly tied to the processes leading to chemical fractionation. 
Resolving this issue will likely require higher quality and more extensive data.

\subsection{Limitations of current solar observations}
\label{limitations}

Currently available observations suitable for abundance analysis suffer from several significant limitations, which are responsible for the relatively modest progress made in setting tight observational constraints on spatial and temporal variation of coronal abundances, essential for the improvement of models of chemical fractionation.  We discuss here some of the current limitations, but we are not aiming at an exhaustive elucidation of all the shortcomings of existing approaches.

Investigations of abundances in the last couple of decades have principally made use of spectral data taken with the EUV spectrometer Hinode/EIS. More recently, EUV spectra observed by Solar Orbiter/SPICE \cite{spice} have become available and have been analyzed also to study abundance anomalies (e.g., \cite{Brooks2024,Varesano2025arXiv250212045V}).
These observations can typically constrain only the abundance of very few elements (e.g., the main abundance diagnostics with Hinode/EIS are based on Si\,{\sc x}/S\,{\sc x} and Ca\,{\sc xiv}/Ar\,{\sc xiv}) and only a few lines are available. Therefore, a major underlying assumption in these studies is that all low-FIP elements vary in unison, and, similarly, that high-FIP elements have constant relative abundances. Several observations both solar and stellar indicate that this assumption is not always warranted (e.g., \cite{Telleschi2005,Wood2010,Drake2011}).
Furthermore, these diagnostics are only probing specific temperatures: e.g., the Ca/Ar diagnostic is only accessible in the hottest portions ($\gtrsim 3$~MK) of the ARs or in flares, as in quiescent conditions the needed lines are generally too weak. Even for the Si/S diagnostic, which is the most used in recent papers investigating coronal abundances, the lines require long exposure times, and are often not very strong, therefore producing quite noisy FIP bias maps (see examples in all the figures in this paper).
Also, when applying the Si/S diagnostic, it is important to keep in mind that S (FIP=10.36~eV) is actually close to the conventional boundary (10~eV) between low-FIP and high-FIP elements, and it is known to behave differently in various environments \cite{Brooks2011,DelZanna2018}. 
SPICE observations provide several abundance diagnostics (e.g., Mg/Ne, S/O, C/O), although often at coarser spatial resolution (2", 4", 6" and 30" slits can be used; along the y direction the resolution is $\sim$ 4", which at the $\sim 0.3$~AU perihelion corresponds to about 1.2" for an instrument at 1~AU) than EIS, but accurate FIP bias diagnostics can be hampered by the difficulty in accounting for the temperature dependence of these diagnostics due to the lack of lines from different ionization stages of the same element covering a sufficient temperature range \cite{Brooks2022}. Additionally, the SPICE abundance diagnostics are from lines mostly formed at transition region temperatures ($\lesssim 0.5$~MK), and there is some indication that the FIP effect might manifest differently at transition region vs coronal temperatures \cite{Laming1995}. 
%The intermittent nature of the Solar Orbiter operations also hinders long time series monitoring a specific target. 

The accuracy of the determination of the abundances is affected by several factors, including the effect of noise (as discussed in the previous paragraph), and the uncertainty in atomic data, which are necessary for all spectral analyses (e.g., \cite{Testa2012,Yu2018,DelZanna2018}). Even when neglecting these effects, the inversion of spectral data to derive the abundances carries other uncertainties, e.g., due to well-known challenges in deriving plasma temperature distributions (see e.g., \cite{Craig1976,Judge1997,Testa2012b,MartinezSykora2026} and references therein), or e.g., to the assumptions made when line ratio methods minimizing the temperature sensitivity are adopted (e.g., \cite{Drake2003,Drake2005,Huenemoerder2009}; a comparison of the two methods is shown e.g., in \cite{Huenemoerder2009,Testa2010}).

In coordinated observations with IRIS and EIS, the spectra from the two instruments have a very low degree of simultaneity (see e.g., Figure~9 in \cite{Testa2016}) due to the characteristics of the instruments, including different spatial and temporal resolution, and even different rastering directions (IRIS in recent years has adopted observing modes W to E like Hinode/EIS, but initially used E to W [left to right] for its raster direction).
Analyses of temporal evolution of FIP bias maps are also hindered by the fact that spatial coverage is often insufficient and inconsistent in time series, even with the same instruments (see e.g., examples in \cite{Testa2023}).

Although no current solar soft X-ray observatory currently provides high-resolution spectral observations, low-resolution soft X-ray spectra are routinely delivered by instruments such as, for instance, Chandrayaan-2 XSM \cite{xsm} and the MinXSS-1 cubesat \cite{minxss}. These instruments provide diagnostics of abundances (e.g., \cite{Suarez2023,Nama2023}), which are, however, limited by the disk-integrated nature and low spectral resolution characterizing their spectra. Rocket experiments such as MaGIXS \cite{magixs} demonstrate the potential for overlapping spatial-spectral data (overlappograms) to enable wide-field imaging soft X-ray spectroscopy at high spectral resolution,  suitable for abundance diagnostics (e.g., \cite{Mondal2025}).

\section{Conclusions and Future Perspectives}
\label{conclusions}

Spectral observations of the solar corona unveiled the peculiar FIP effect, i.e., low-FIP elements are often enhanced in the solar outer atmosphere. During the past couple of decades, spatially and temporally resolved EUV coronal spectra have uncovered a complex scenario where the degree of enhancement of low-FIP elements in the corona seem to vary substantially among different structures depending on several factors, including activity level, magnetic topology, and occasionally showing even an inverse FIP effect (see e.g., \cite{Baker2015,JMS2025} in this topical issue, and references therein), and often showing high temporal variability (e.g.,\cite{Brooks2025} in this volume). 
Stellar observations have shown that the fractionation processes are not unique to the Sun but are a fundamental property of magnetically heated atmospheres, and provide a powerful comparative laboratory for testing fractionation models, constraining chromospheric physics across a wide range of magnetic and energetic regimes (e.g., \cite{testa2015}).

In this paper we focused on reviewing the first attempts made recently at tracing signatures of the coronal abundance anomalies to chromospheric properties. The relation of the coronal FIP bias with plasma properties in the lower atmosphere, where the fractionation is expected to happen, would provide valuable constraints on the fractionation mechanisms.
The observations suggest a possible enhancement of chromospheric microturbulence in outflow regions, that are typically characterized by high FIP bias. Additionally, in some high FIP bias areas the transition region non-thermal broadening also appears enhanced. However, these enhancements do not appear consistently for all high FIP bias regions, such as for instance for regions close to sunspots \cite{Testa2023,Long2024}.
While keeping in mind the caveat that these results are not conclusive, at least in part because of the limitations in the observations (which we discussed in some detail in 
Sec.~\ref{fip_chromo}\ref{limitations}) and the paucity of observations suitable for investigating this issue, we can speculate on the possible implications of these findings. These results are puzzling and raise several questions, such as for instance:
Is the fractionation mechanism very different in regions with significantly dissimilar magnetic properties (e.g., AR outflows vs sunspots vs moss)? Why is the FIP effect close to sunspots often very different from other coronal structures? How spatially and temporally structured are the fractionation processes?

A deeper understanding of the nature of the FIP effect and its broad manifestations will certainly require both advancing fractionation models and deriving much tighter constraints from the observations, across different atmospheric layers.

Currently available models \cite{Laming2015,MartinezSykora2023} have made significant strides in building a theoretical framework for the interpretation of the observation. However, their limitations need to be addressed: the semi-analytical 1D models of \cite{Laming2009,Laming2015} assume a static atmosphere, prescribe the wave spectra, and cannot capture the chromospheric complexity, but, given the significant number of free parameters, appear to be able to predict expected abundance patterns; on the other hand, the time-dependent multi-fluid multi-dimensional MHD models of \cite{MartinezSykora2023} can simulate realistic chromospheric wave fields, currents, and ion-neutral coupling, and also investigate chemical fractionation driven by other physics processes, e.g., reconnection \cite{Wargnier2023ApJ...946..115W,Wargnier2025AA...695A.262W}, but they are not advanced enough yet to produce quantitative, observable abundance predictions.

The scarcity of observations suitable to link the coronal abundances to the underlying conditions in the lower atmosphere is partly due to the difficulty of obtaining high quality coordinated observations with different instruments probing the different atmospheric layers. The planned EUV High-Throughput Spectroscopic Telescope (EUVST \cite{EUVST}) mission, thanks to its broad wavelength coverage and high throughput, would overcome the limitations of current observations by providing diagnostics of chemical fractionation (e.g., with strong lines of S\,{\sc v}, O\,{\sc v}, O\,{\sc vi}, Ne\,{\sc vii}, Ne\,{\sc viii}, and Fe\,{\sc viii}-Fe\,{\sc xxi}) and simultaneously observing the atmosphere from the chromosphere to the corona, at high spatial and temporal resolution, and high signal-to-noise. While waiting for EUVST, targeted coordinated observations of the lower atmosphere with IRIS, spectropolarimetric observations to constrain the Alfv{'e}n wave spectra (e.g., with DKIST), and transition region and coronal data with Hinode/EIS and SolO/SPICE to derive FIP bias maps, and PSP/Solar Orbiter data to establish the connectivity to the solar wind, will help track chromospheric wave/turbulence proxies and overlying FIP bias along mapped field lines. 

Understanding the nature and causes of the observed abundance anomalies is of fundamental importance, and solar observations can uniquely constrain models that will help us interpret the physical processes at work, and also to interpret unresolved observations of other solar-like stars.

\ack{
 We thank the Royal Society for its funding and support throughout the organization of the Theo Murphy meeting "Solar Abundances in Space and Time" and the production of this special issue. We also thank the referees for their comments which have helped improve the paper.
 The work of PT was supported by NASA contract NNM07AB07C (Hinode/XRT) to the Smithsonian Astrophysical Observatory, contract 8100002705 (IRIS) to the Smithsonian Astrophysical Observatory, contract SP02H1701R (AIA) to the Smithsonian Astrophysical Observatory, NASA Heliophysics Guest Investigator grant 80NSSC21K0737, and NASA Heliophysics Supporting Research grant 80NSSC21K1684. JMS  was supported NNG09FA40C (IRIS) and 80GSFC21C0011 (MUSE), NASA grant 80NSSC26K0018. BDP was supported by NASA contract NNG09FA40C (IRIS). D.B.\ is funded under Solar Orbiter EUI Operations grant number ST/X002012/1 and Hinode Ops Continuation 2022-25 grant number ST/X002063/1. The work of D.H.B.\ was performed under contract to the Naval Research Laboratory and was funded by the NASA Hinode project.}
%%%%%%%%%% Insert bibliography here %%%%%%%%%%%%%%

\bibliographystyle{RS}

\end{document}